\documentclass[fleqn,usenatbib]{mnras}

\usepackage{newtxtext,newtxmath}
\usepackage[T1]{fontenc}

\DeclareRobustCommand{\VAN}[3]{#2}
\let\VANthebibliography\thebibliography
\def\thebibliography{\DeclareRobustCommand{\VAN}[3]{##3}\VANthebibliography}

\usepackage{graphicx}	% Including figure files
\usepackage{amsmath}	% Advanced maths commands
\title[NEOROCKS: surface properties of small NEOs]{NEOROCKS project: surface properties of small near-Earth asteroids}

\author[T.~Hromakina et al.]{
T.~Hromakina,$^{1,2}$\thanks{E-mail: tetiana.hromakina@obspm.fr}
M.~Birlan,$^{3,4}$
M.~A.~Barucci,$^{1}$
M.~Fulchignoni,$^{1}$
F.~Colas,$^{3}$
S.~Fornasier,$^{1,5}$
F.~Merlin,$^{1}$
\newauthor
A.~Sonka,$^{4}$
S.~Anghel,$^{3,4,6}$
G.~Poggiali,$^{1}$
I.~Belskaya,$^{1,2}$
D.~Perna$^{7}$
E.~Dotto$^{7}$
and the NEOROCKS team \thanks{The NEOROCKS team: E. Dotto, M. Banaszkiewicz, S. Banchi, M.A. Barucci, F. Bernardi, M. Birlan, B. Carry, A. Cellino, J. De Leon, M. Lazzarin, E. Mazzotta Epifani, A. Mediavilla, J. Nomen Torres, D. Perna, E. Perozzi, P. Pravec, C. Snodgrass, C. Teodorescu, S. Anghel, A. Bertolucci, F. Calderini, F. Colas, A. Del Vigna, A. Dell’Oro, A. Di Cecco, L. Dimare, P. Fatka, S. Fornasier, E. Frattin, P. Frosini, M. Fulchignoni, R. Gabryszewski, M. Giardino, A. Giunta, T. Hromakina, J. Huntingford, S. Ieva, J.P. Kotlarz, F. La Forgia, J. Licandro, H. Medeiros, F. Merlin, F. Pinna, G. Polenta, M. Popescu, A. Rozek, P. Scheirich, A. Sergeyev, A. Sonka, G.B. Valsecchi, P. Wajer, A. Zinzi.}
\\
\\
$^{1}$LESIA, Universit{\'e} Paris Cit{\'e}, Observatoire de Paris, Universit{\'e} PSL, Sorbonne Université, CNRS, F-92190 MEUDON, France\\
$^{2}$V.~N.~Karazin Kharkiv National University, 4 Svobody Sq., Kharkiv, 61022, Ukraine\\
$^{3}$IMCCE, Observatoire de Paris, CNRS UMRO 8028, PSL Research University, 77 av Denfert Rochereau, 75014, Paris Cedex, France\\
$^{4}$Astronomical Institute of the Romanian Academy, 5 Cutitul de Argint, 040557, sector 4, Bucharest, Romania\\
$^{5}$Institut Universitaire de France (IUF), 1 rue Descartes, 75231 PARIS CEDEX 05\\
$^{6}$Faculty of Physics, University of Bucharest, 405, Atomistilor Street, 077125 Magurele, Ilfov, Romania\\
$^{7}$INAF – Osservatorio Astronomico di Roma, Via Frascati 33, 00078 Monte Porzio Catone, Italy\\
}

\date{Accepted XXX. Received YYY; in original form ZZZ}

\pubyear{2022}

\begin{document}
\label{firstpage}
\pagerange{\pageref{firstpage}--\pageref{lastpage}}
\maketitle

\begin{abstract}
We present new results of the observing program which is a part of the NEOROCKS project aimed to improve knowledge on physical properties of near-Earth Objects (NEOs) for planetary defense.
Photometric observations were performed using the 1.2m telescope at the Haute-Provence observatory (France) in the BVRI filters of the Johnson-Cousins photometric systems between June 2021 and April 2022. We obtained new surface colors for 42 NEOs. Based on the measured colors we classified 20 objects as S-complex, 9 as C-complex, 9 as X-complex, 2 as D-type, one object as V-type, and one object remained unclassified. For all the observed objects we estimated their absolute magnitudes and diameters. Combining these new observations with the previously acquired data within the NEOROCKS project extended our dataset to 93 objects. The majority of objects in the dataset with diameters D<500m belongs to a group of silicate bodies, which could be related to observational bias. Based on MOID and $\Delta$V values we selected 14 objects that could be accessible by a spacecraft. Notably, we find D-type asteroid (163014) 2001 UA5 and A-type asteroid 2017 SE19 to be of particular interest as possible space mission targets.
\end{abstract}


%%%%%%%%%%%%%%%%%%%%%%%%%%%%%%%%%%%%%%%%%%%%%%%%%%

%%%%%%%%%%%%%%%%% BODY OF PAPER %%%%%%%%%%%%%%%%%%

\section{Introduction}

A population of near-Earth objects (NEOs) is believed to be brought into their current location from the asteroid belt due to the Yarkovsky effect and gravitational interactions with Jupiter and Saturn and defined by perihelion distance $q$<1.3~au \citep{Granvik2017, Bottke2006, Wisdom1983}.
Such proximity to our planet gives us a chance to study these asteroids down to several meters in diameter, whereas main belt asteroids could be reached only down to hundreds of meters by the means of ground-based observations. 
Investigation of NEOs provides us with an opportunity to better understand the evolution of the solar system and the processes that took place at the early stages of its formation. 
Notably, NEOs are also a threat to humanity due to potential collision hazard \citep[e.g.,][]{Perna2013,Perna2016}. In particular, NEOs with minimum orbital intersection distance MOID<0.05~au and absolute magnitude H<22 are classified as potentially hazardous asteroids (PHAs). 

The number of discovered NEOs is constantly growing, but the number of NEOs with known physical properties is very small. According to the Light Curve Database \citep[][updated 14 December, 2021]{Warner2009}, rotational period and/or surface color is known for 2117 objects, which is only about 7\% of the discovered NEOs.
In addition, the majority of these objects are large NEOs. The rate at which new NEOs are being discovered only increases this gap. Thus, there is a need for dedicated surveys that would focus on investigating the physical properties of NEOs, particularly the smallest members of the population.

The whole variety of compositional classes is found among NEOs, with majority of the population (from 40 to 70\%) represented by silicate S-complex asteroids \citep{Devogele2019, Ieva2018, Ieva2020, Lin2018, Binzel2019, Popescu2019, Perna2018}. However, the recent work by \citet{Marsset2022} that presented a debiased distribution of NEOs suggests that the low-albedo carbonaceous asteroids might be underrepresented, and the amount of silicate and carbonaceous asteroids is nearly equal.

Here we present the results of new observations conducted within the framework of the NEOROCKS project, which aims to characterize small NEOs by obtaining their surface colors and classify them into taxonomic classes. 

In Section~\ref{sec_obs} we describe the observations and main results. Section~\ref{sec_taxon} presents taxonomic classification of the observed NEOs. We present the analysis and discuss the results in Section~\ref{sec_disc}. Finally, the summary and conclusions of this work are presented in Section~\ref{concl}.

\section{Observations and results}
\label{sec_obs}

New photometric data presented in this work were obtained during several observational runs between June 2021 and April 2022 using the 1.2m telescope at the Haute-Provence observatory (OHP), located in France. The telescope is equipped with a 2048$\times$2048 Andor Ikon L 936 CCD camera that has a field-of-view of 13.1'$\times$13.1'. A 2$\times$2 binning that was applied throughout all observing runs gave a 0.77''/pxl scale.

The selection of the targets was similar to the previous work \citep{Hromakina2021}: we prioritised objects with higher absolute magnitude (i.e. smaller objects) and PHAs. Additionally, newly discovered NEOs were also added to the priority list, because the discovery apparition is followed by a small window when the object is available for characterisation \citep[e.g., ][]{Galache2015}. 

Observations were obtained in B-V-R-I filters of the Johnson-Cousins photometric system. Each object was observed for about an hour, which, depending on the exposure time from 30 seconds to 3 minutes, resulted in 3-15 images in each filter. The images in each filter were taken sequentially in order to minimize possible magnitude variation due to the rotation of an object.

Data reduction was done following the same approach that was described in \citet{Hromakina2021}, which includes bias subtraction and flat-field correction. The instrumental magnitudes were measured for the target and field stars using aperture photometry. Then, absolute calibration was done using magnitudes of field stars in the Pan-STARRS catalog. To transform star magnitudes from the Sloan photometric system of the Pan-STARRS1 catalog into the Johnson-Cousins system we used transformational equations presented in \citet{Kostov2018}.

We observed 42 small NEOs, about a quarter of which are recently discovered objects. Five objects observed during the April 2022 run were observed as part of the Rapid Response Experiment (Perna et al., in preparation), which is another objective of the NEOROCKS programme. Nearly 40\% of the observed NEOs (16 out of 42) are classified as PHAs. Table~\ref{table_colors} contains the photometry results, estimated absolute magnitudes and diameters, and observational circumstances of our data. 
The orbital elements of the observed asteroids are presented in Table~\ref{table1} of the Appendix~\ref{annex-table}. 

\begin{figure}
    \centering
    \includegraphics[width=0.9\columnwidth]{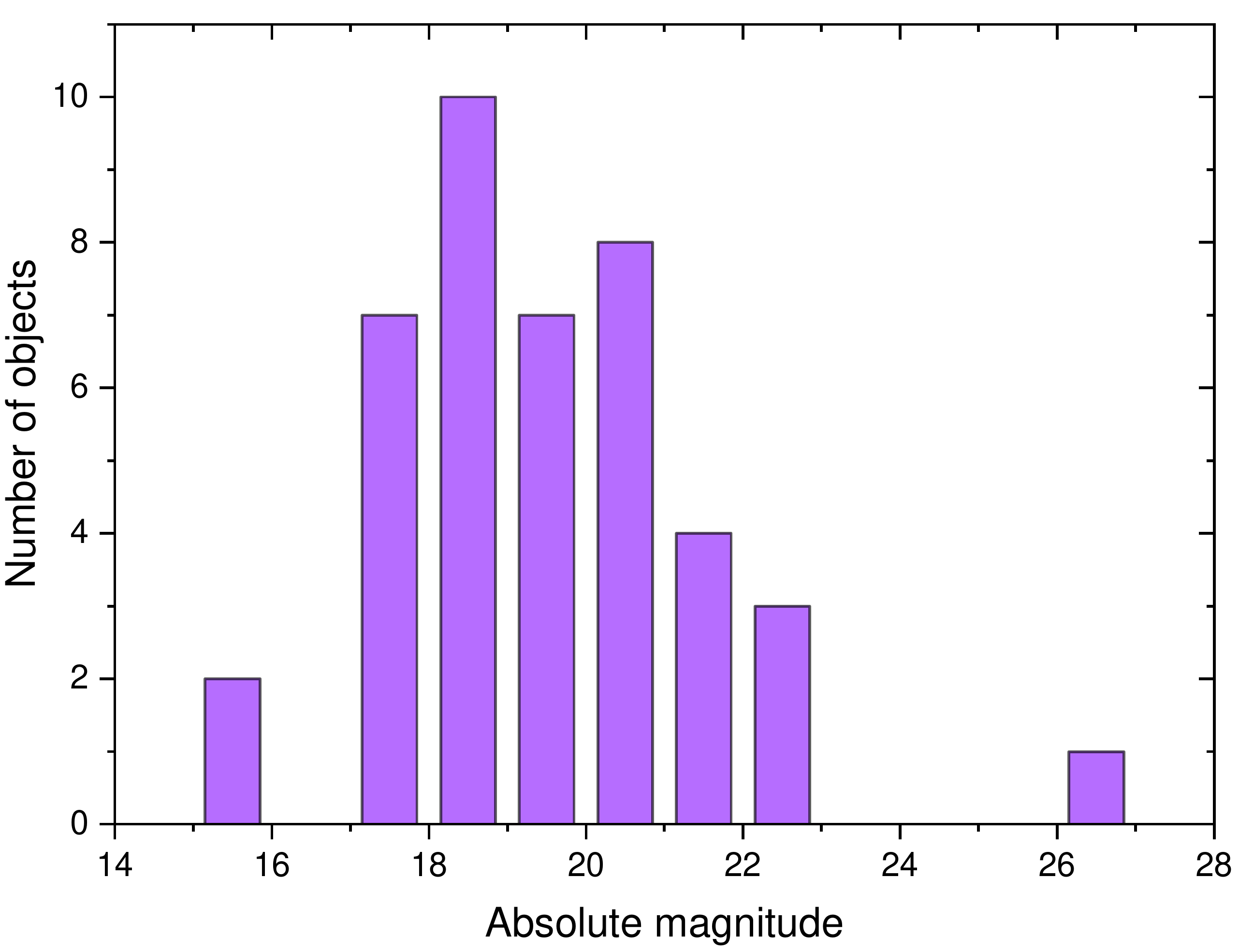}
    \caption{Absolute magnitude distribution of the observed NEOs.}
    \label{H_hist}
\end{figure}

Absolute magnitudes (H) given in Table~\ref{table_colors} were calculated using our measurements of V magnitudes via the online tools for H,G$_{1}$, G$_{2}$ photometric system\footnote{https://wiki.helsinki.fi/display/PSR/HG1G2+tools}. It was shown that the one-parameter version of the H, G$_{1}$, G$_{2}$ function can give reliable results even in the case of only one magnitude measurement \citep{Penttila2016}. We compared our absolute magnitudes with the values presented in the Minor Planetary Center (MPC) and found generally good agreement. For three asteroids 2010 WQ7, 2022 BQ3, and 2022 DC5 the discrepancies exceed 0.5 mag. Further observations are needed to find possible reasons for these discrepancies. Absolute magnitude distribution of the sample is shown in Fig.~\ref{H_hist}. Almost all of the observed NEOs fall into the 17-23 mag range, with the largest absolute magnitude reaching H$\sim$26. 

To estimate the diameters of the observed objects we used the obtained absolute magnitudes and albedos of the respective taxonomic classes from \citet{Ryan2010}. For the unclassified object 2022 GY2 we assumed S-complex taxonomic class. For 4 out of 42 NEOs in our sample the diameters were measured by the NEOWISE survey \citep{Mainzer2019}. Our estimations are well-consistent with the measured values within the uncertainties. In Table~\ref{table_colors} we provide the NEOWISE diameters for four asteroids. 

Table~\ref{table_colors} also contains the results of the taxonomic classification that was performed in this work and described in the next section. 

\begin{table*}
\scriptsize
    \centering
    \caption{Observational circumstances and results.}
    \label{table_colors}
        \begin{tabular}{lllllllllllll}
\hline
\hline
&Object&Date&r, au& $\Delta$, au &$\alpha$, deg &V mag&B-V&V-R& V-I & Taxon& H & D, m \\
\hline
1&(12923) Zephyr&27/11/2021&1.408&0.618&37.1& 17.21$\pm$0.02 &0.94$\pm$0.04&0.45$\pm$0.03&0.70$\pm$0.04&Q/S & 16.1   &2060$\pm$10*\\
2&(87024) 2000 JS66&09/06/2021&1.063&0.166&69.1& 17.72$\pm$0.02 &0.71$\pm$0.02&0.37$\pm$0.02&0.83$\pm$0.04&X&  19.0    &670-1080\\
3&(89958) 2002 LY45&11/06/2021&1.305&0.362&31.9 &16.94$\pm$0.02 &0.75$\pm$0.04&0.47$\pm$0.02&0.75$\pm$0.02&Q/S& 17.3   &1100-1330 \\
4&(138971) 2001 CB21&28/01/2022&1.165&0.251&39.6  &17.43$\pm$0.03 &0.92$\pm$0.04&0.39$\pm$0.03&0.82$\pm$0.04&X&  18.4  & 940-1190\\
&&29/01/2022&1.160&0.242&39.5                       &17.24$\pm$0.03&0.84$\pm$0.04&0.43$\pm$0.02&0.79$\pm$0.03&\\
5&(140158) 2001 SX169&09/06/2021&1.097&0.128&47.5 &15.89$\pm$0.01 &0.76$\pm$0.03&0.46$\pm$0.02&0.82$\pm$0.02&Xe/X& 18.4  &566$\pm$10* \\
6&(143649) 2003 QQ47&02/10/2021&1.068&0.173&63.1  &16.12$\pm$0.02 &0.90$\pm$0.03&0.52$\pm$0.02&0.86$\pm$0.02&S& 17.5     &950-1160\\
7&(162913) 2001 MT18&09/06/2021&1.017&0.206&83.8 &18.06$\pm$0.02 &&0.55$\pm$0.02&1.05$\pm$0.03&D&   18.5    &940-1310\\
8&(163692) 2003 CY18&05/04/2022 &1.307&0.355&26.7  &18.03$\pm$0.02  &0.84$\pm$0.03&0.49$\pm$0.02&0.94$\pm$0.03&S& 18.5     &630-760\\
9&(318160) 2004 QZ2&27/01/2022&1.389&0.407&5.7   &17.46$\pm$0.02 &0.81$\pm$0.02&0.47$\pm$0.02&0.77$\pm$0.03&Sq/S& 18.2    &690-840\\
10&(363027) 1998 ST27&01/10/2021&1.152&0.194&36.1  &17.72$\pm$0.02 &0.78$\pm$0.03&0.42$\pm$0.02&0.75$\pm$0.03&C& 19.6   &578$\pm$228*\\
&&02/10/2021&1.146&0.192&37.5                       &17.78$\pm$0.02 &0.73$\pm$0.02&0.39$\pm$0.02&0.72$\pm$0.03&\\
11&(363599) 2004 FG11& 05/04/2022&1.082&0.102&35.7  &17.57$\pm$0.02  &  0.76$\pm$0.02&0.43$\pm$0.02&0.74$\pm$0.02&Xe/X& 21.2  & 152$\pm$3*\\
&&06/04/2022&   1.069&0.090&39.5                    &17.18$\pm$0.02& 0.77$\pm$0.02&0.43$\pm$0.02&0.71$\pm$0.02\\
12&(366746) 2004 LJ&10/06/2021&1.112&0.142&44.5   &17.80$\pm$0.02 &&0.51$\pm$0.03&0.76$\pm$0.03&Sq/S& 20.15  &  260-320\\
13&(374855) 2006 VQ13&26/11/2021&1.138&0.176&28.6 &17.85$\pm$0.02 &0.71$\pm$0.04&0.44$\pm$0.03&0.82$\pm$0.10&X&     20.1  &370-590\\
&&27/11/2021&1.145&0.182&27.5                       &17.89$\pm$0.03 &&0.50$\pm$0.03&0.87$\pm$0.03&\\
14&(388945) 2008 TZ3&05/04/2022&1.185&0.203&22.6  &18.33$\pm$0.03 &0.73$\pm$0.03&0.41$\pm$0.03&0.68$\pm$0.04&Cg/C& 20.3    &370-480\\
15&(410195) 2007 RT147&28/01/2022&1.361&0.383&9.2  &17.32$\pm$0.02 &0.75$\pm$0.04&0.45$\pm$0.03&0.66$\pm$0.04&Q/S&  18.2  &580-700\\
16&(415029) 2011 UL21&10/06/2021&1.181&0.540&59.1  &16.90$\pm$0.02 &0.82$\pm$0.04&0.47$\pm$0.02&0.86$\pm$0.03&Q/S&  15.7   & 1900-2310\\
&&11/06/2021&1.191&0.542&58.2                       &16.90$\pm$0.02 &0.78$\pm$0.04&0.49$\pm$0.02&0.88$\pm$0.03&\\
17&(450263) 2003 WD158&08/06/2021&1.007&0.072&94.6&16.91$\pm$0.02 &&0.52$\pm$0.03&0.90$\pm$0.04&S& 19.4  & 460-550 \\
&&10/06/2021&1.014&0.066&89.1                       &16.77$\pm$0.02 &0.84$\pm$0.03&0.48$\pm$0.02&0.86$\pm$0.03&\\
18&(475665) 2006 VY13 &06/04/2022&1.338&0.456&35.3 &17.72$\pm$0.02& &0.35$\pm$0.03&0.65$\pm$0.03&C& 17.4   & 1630-2070  \\
19&(491567) 2012 RG3&10/06/2021&1.242&0.382&46.1   &18.34$\pm$0.03 &&0.41$\pm$0.03&0.92$\pm$0.03&D&  18.3   & 1030-1430\\
20&(495615) 2015 PQ291&09/06/2021&1.215&0.334&46.5 &17.37$\pm$0.02 &0.83$\pm$0.03&0.47$\pm$0.02&0.88$\pm$0.03&S&    17.6    & 790-960\\
21&(506459) 2002 AL14&29/01/2022&1.127&0.152&19.8  &15.10$\pm$0.01 &0.88$\pm$0.02&0.50$\pm$0.02&0.92$\pm$0.02&S&    18.0    & 760-920\\
22&(516396) 2000 WY28&26/11/2021&1.165&0.180&7.0   &17.88$\pm$0.02 &0.90$\pm$0.05&0.36$\pm$0.05&0.67$\pm$0.05&C&   20.8     & 430-550\\
23&(613291) 2005 YX128&28/01/2022&1.072&0.176&56.0 &17.35$\pm$0.02 &0.80$\pm$0.03&0.42$\pm$0.02&0.72$\pm$0.03&C&  18.9    & 710-910\\
24&(613403) 2006 GB&04/04/2022&1.056&0.107&55.8    &17.39$\pm$0.02&0.77$\pm$0.02&0.40$\pm$0.02&0.77$\pm$0.03&C&     20.1   & 390-500\\
25&2002 TP69&26/11/2021&1.041&0.057&17.7      &16.81$\pm$0.01 &0.84$\pm$0.03&0.50$\pm$0.02&0.93$\pm$0.02&S&  22.0    &120-150 \\
&&27/11/2021&1.043&0.059&16.3                  &16.85$\pm$0.02 &0.87$\pm$0.03&0.51$\pm$0.02&0.92$\pm$0.02&\\
26&2009 CC3&05/04/2022&1.045&0.143&68.1     &17.38$\pm$0.02&0.96$\pm$0.04&0.52$\pm$0.02&0.77$\pm$0.02&S&   19.1   & 460-550 \\
27&2010 TV149& 04/04/2022&1.079&0.177&60.3   &17.08$\pm$0.02&0.78$\pm$0.03&0.42$\pm$0.02&0.79$\pm$0.03&X&   18.5  & 670-1080\\
28&2010 WQ7&28/01/2022&1.130&0.248&48.8        &18.26$\pm$0.02 &0.75$\pm$0.04&0.51$\pm$0.03&0.88$\pm$0.03&S/S&  19.2  &630-760\\
29&2011 YQ10&26/11/2021&1.308&0.324&6.9        &17.83$\pm$0.02 &0.72$\pm$0.02&0.49$\pm$0.02&0.63$\pm$0.02&Q/S&  19.1    &440-530\\
30&2017 UW42&27/01/2022&1.268&0.336&28.2       &17.34$\pm$0.01 &0.87$\pm$0.02&0.50$\pm$0.02&0.91$\pm$0.02&S&    17.9    &760-920 \\
31&2018 CW13&28/01/2022&1.177&0.193&5.0         &17.03$\pm$0.02 &0.80$\pm$0.03&0.43$\pm$0.02&0.72$\pm$0.03&Cg/C& 19.9    &450-570\\
32&2021 JQ24&27/11/2021&1.378&0.420&18.0     &17.53$\pm$0.02 &0.67$\pm$0.03&0.40$\pm$0.02&0.73$\pm$0.03&Xc/X&   17.8    &1010-1630\\
33&2021 LN3&11/06/2021&1.103&0.093&19.2       &17.22$\pm$0.02 &&0.46$\pm$0.03&0.73$\pm$0.05&Q/S&               21.2   &150-180\\
34&2021 SR41**& 06/04/2022&1.144&0.275&52.5       &19.14$\pm$0.03&0.82$\pm$0.03&0.39$\pm$0.03&0.72$\pm$0.03&C&   19.7  &470-600\\
35&2021 VM25**& 08/04/2022&1.043&0.073&53.1     &16.71$\pm$0.02&0.71$\pm$0.03&0.49$\pm$0.02&0.86$\pm$0.03&S& 20.4  &360-500  \\
36&2021 WX3&04/04/2022&1.116&0.132&27.1        &17.48$\pm$0.02&0.71$\pm$0.02&0.38$\pm$0.02&0.67$\pm$0.02&C&    20.4  &   360-450 \\
37&2022 BH3&29/01/2022&0.998&0.016&31.6         &18.57$\pm$0.02&0.91$\pm$0.03&0.51$\pm$0.02&&S& 26.3     &10-20\\
38&2022 BK&27/01/2022&1.037&0.057&23.4       &17.66$\pm$0.02&&0.43$\pm$0.02&0.82$\pm$0.03&X&               22.7       &110-170\\
39&2022 BQ3& 05/04/2022&1.147&0.218&43.5       &17.72$\pm$0.02&1.00$\pm$0.03&0.5$\pm$0.02&0.64$\pm$0.03&V&  19.1      &240-280\\
40&2022 DC5**& 06/04/2022&1.039&0.068&53.6       &18.62$\pm$0.02&&0.43$\pm$0.02&0.70$\pm$0.03&X&    22.4      &80-90\\
41&2022 GC1**& 04/04/2022&1.112&0.117&15.9      &18.94$\pm$0.02&&0.52$\pm$0.02&0.87$\pm$0.03&S&     22.5      &80-100 \\
42&2022 GY2**& 08/04/2022&1.051&0.059&31.8        &17.04$\pm$0.03&&0.38$\pm$0.03&& &  21.8&    120-190\\
    \hline    \end{tabular}
    \begin{flushleft}
$^*$ Measured by NEOWISE survey \citep{Mainzer2019}. \\
$^{**}$These objects were observed as part of the Rapid Response Experiment (Perna et al., in preparation).\\
    \end{flushleft}  
\end{table*}

\section{Taxonomy}
\label{sec_taxon}

Taxonomic classification was done using the obtained surface colors B-V, V-R, and V-I, which were transformed into reflectances using the following equation:  
\begin{equation}
R(\lambda)=10^{-0.4[(M_\lambda-M_V)-(M_\lambda-M_V)_\odot]},
\end{equation}
where $(M_\lambda-M_V)$ and $(M_\lambda-M_V)_\odot$ are the colors of the object and the Sun at the wavelength $\lambda$, respectively. 
Solar colors were taken from \citet{Holmberg2006}. The obtained reflectance was normalized at the central wavelength of the V-filter.
Then, using the M4AST service \citep{Popescu2012, Birlan2016}, the resulting spectra were compared to the mean spectra of the taxonomic classes in DeMeo classification \citep{DeMeo2009}. In the case when the object was observed for two nights the average surface colors were used for taxonomic classification. One asteroid from the dataset, 2022 GY2, was not classified because only V-R color was measured. 
As our taxonomic classification is based on only three or four data points, it is not nearly as precise as spectral data. Thus, in Table~\ref{table_colors} we indicate the major taxonomic classes, such as S-complex, C-complex, X-complex, A-type, D-type, and V-type. However, in the case of an unambiguous match to a specific sub-type according to Bus-DeMeo classification, we indicate both the sub-type and the major type. Some asteroids in our sample revealed similarity with Q-type, first defined by \citep[][]{Tholen1984} for a near-Earth asteroid and considered as an end member class in \citet{DeMeo2009}. In our further analysis we consider Q-type as a part of the S-complex.
Figure~\ref{indiv_plots} in Appendix \ref{annex-figure} shows individual reflectance spectra of asteroids together with the mean spectra of the best-matching taxonomic class.

\begin{figure}
    \centering
    \includegraphics[width=0.9\columnwidth]{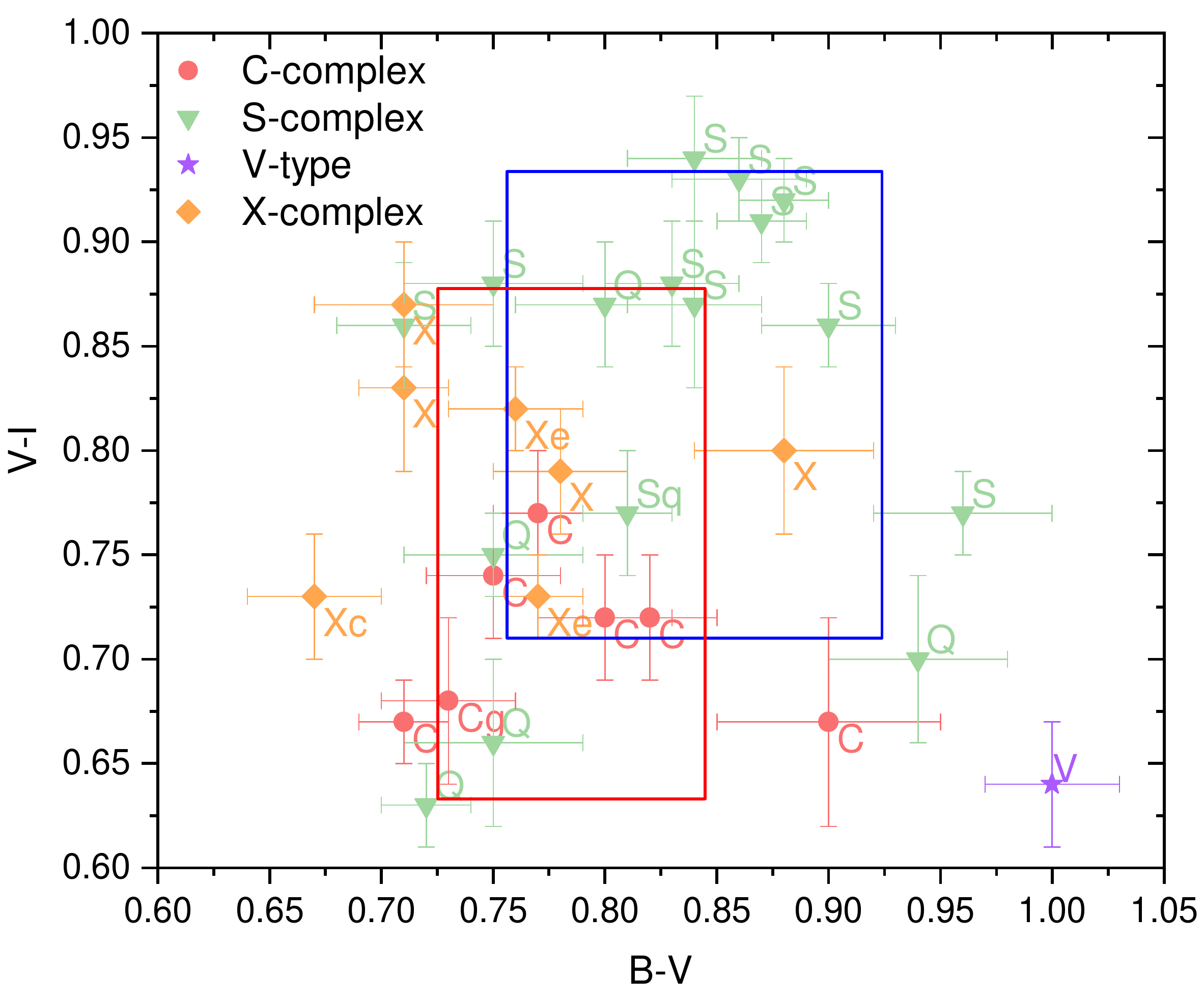}
        \qquad
    \includegraphics[width=0.9\columnwidth]{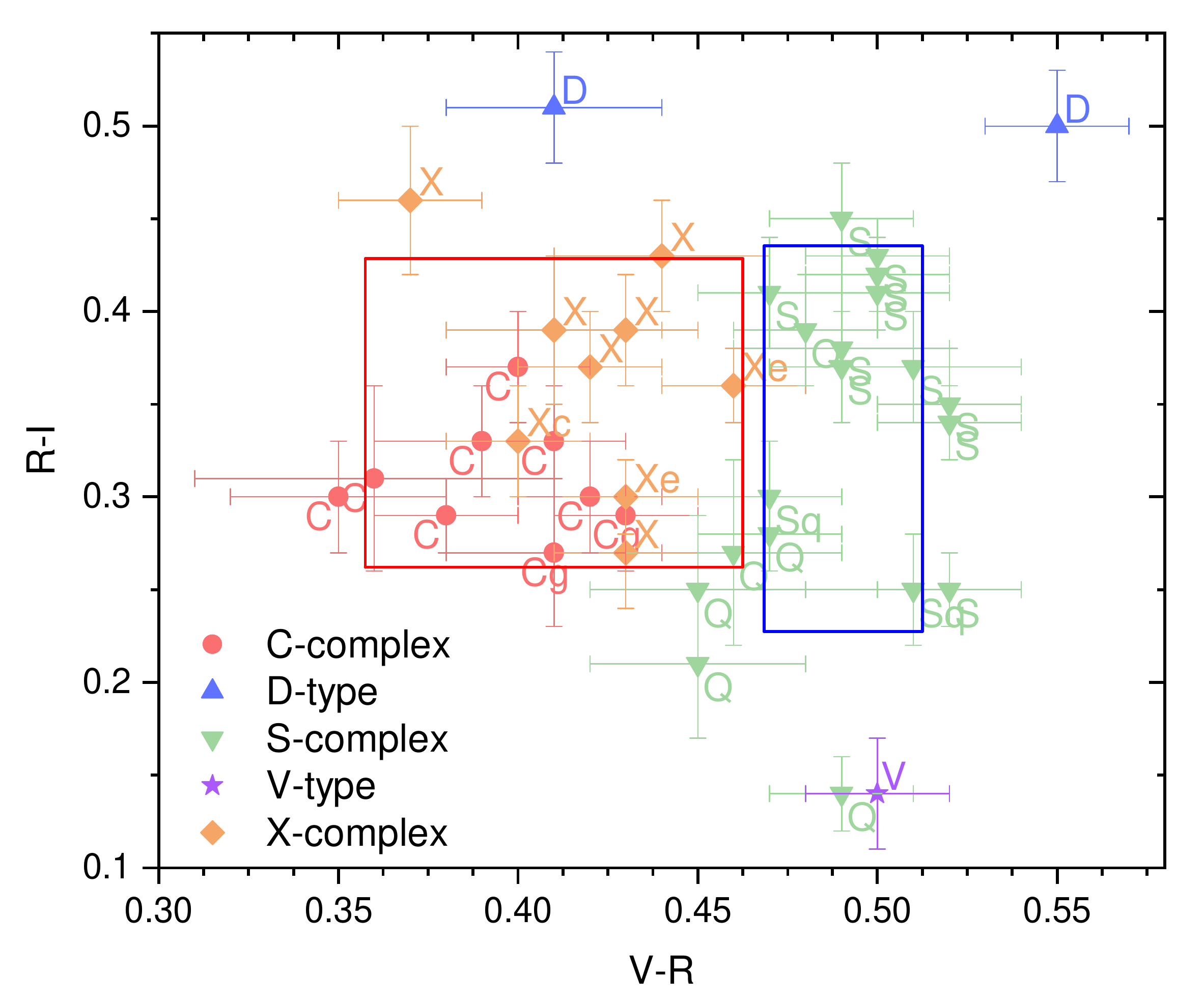}
    \caption{Color-color diagrams showing main taxonomic classes for our dataset. The boxes correspond to the 1$\sigma$ deviation from the mean color values for the group of silicate (blue box) and carbonaceous (red box) objects. }
    \label{color-color}
\end{figure}

We checked our classification using color-color plots (Figure~\ref{color-color}). Boxes in the figures represent 1$\sigma$ deviation of silicate and carbonaceous objects (see sect.~\ref{sec_disc}).
Some of the asteroids in the color-color diagrams seem to be "outliers", so we looked more closely into these objects. 
Asteroid (162913) 2001 MT18 is classified as D-type and has very red surface colors. Taking into account that this NEO was observed at a large phase angle (83.8$^\circ$), its very red colors may be a result of phase reddening \citep[e.g.,][]{Schroder2014, Perna2018}. Other S-complex "outliers" are classified as Q-types, fresh-surfaced asteroids with more neutral slopes compared to S-type asteroids.

One more way to establish the reliability of the taxonomic classification is to compare the obtained results for the objects that were already classified based on more informative spectral data. Only two objects in this work, namely (12923) Zephyr and (506459) 2002 AL14 were previously classified by using spectral data as S-complex and L-type, respectively \citep{Binzel2004}. And in this work both objects were classified into the S-complex class. 

Additionally, geometric albedo values are available for four NEOs. Among them, (12923) Zephyr, (140158) 2001 SX169, and (363599) 2004 FG11 that have moderate albedo compatible with the classification of these objects into S- and X-complexes. Similarly, NEO (363027) 1998 ST27 has a low albedo and is classified into C-complex in this work.

\section{Data analysis and discussion}
\label{sec_disc}

For the objects in our dataset, we calculated the average colors for each taxonomic class and compared them with those reported in the literature. Namely we use colors from \citet{DeMeo2009}, that were derived from the spectral data, and colors from \citet{Ieva2018} and \citet{Lin2018}, that were obtained from the broadband photometric observations (see Table~\ref{table_mean_colors}). We found that the average colors obtained in this work are in general agreement with the literature values. We do not include A- and V-types in Table~\ref{table_mean_colors} as we have only a few asteroids classified into these taxons.

\begin{table*}
\small
    \centering
    \caption{Average colors obtained in this work for each taxonomic class in comparison to the literature values from \citet{DeMeo2009}, \citet{Ieva2018}, and \citet{Lin2018}.}
    \label{table_mean_colors}
        \begin{tabular}{lllllllll}
 
\hline\hline        
Taxonomic class&B-R&B-R $_{Lin18}$ &B-R $_{Ieva18}$ &B-R $_{DeMeo09}$ &V-I &V-I$_{Lin18}$ &V-I$_{Ieva18}$ &V-I$_{DeMeo09}$ \\
\hline
C&1.18$\pm$0.09&1.06&1.07$\pm$0.12&1.09$\pm$0.11&0.69$\pm$0.06&0.67&0.67$\pm$0.07&0.69$\pm$0.10\\
D&1.18$\pm$0.05&1.19&1.21+0.06&1.18$\pm$0.07&0.92$\pm$0.08&0.87&0.89$\pm$0.03&0.89$\pm$0.07\\
S&1.32$\pm$0.07&1.27&1.32$\pm$0.07&1.29$\pm$0.14&0.83$\pm$0.11&0.81&0.77$\pm$0.09&0.82$\pm$0.14\\
X&1.14$\pm$0.05&1.13&1.19$\pm$0.06&1.13$\pm$0.12&0.80$\pm$0.04&0.79&0.78$\pm$0.05&0.78$\pm$0.06\\
    \hline
    \end{tabular}
\end{table*}

\begin{figure}
    \centering
    \includegraphics[width=0.95\columnwidth]{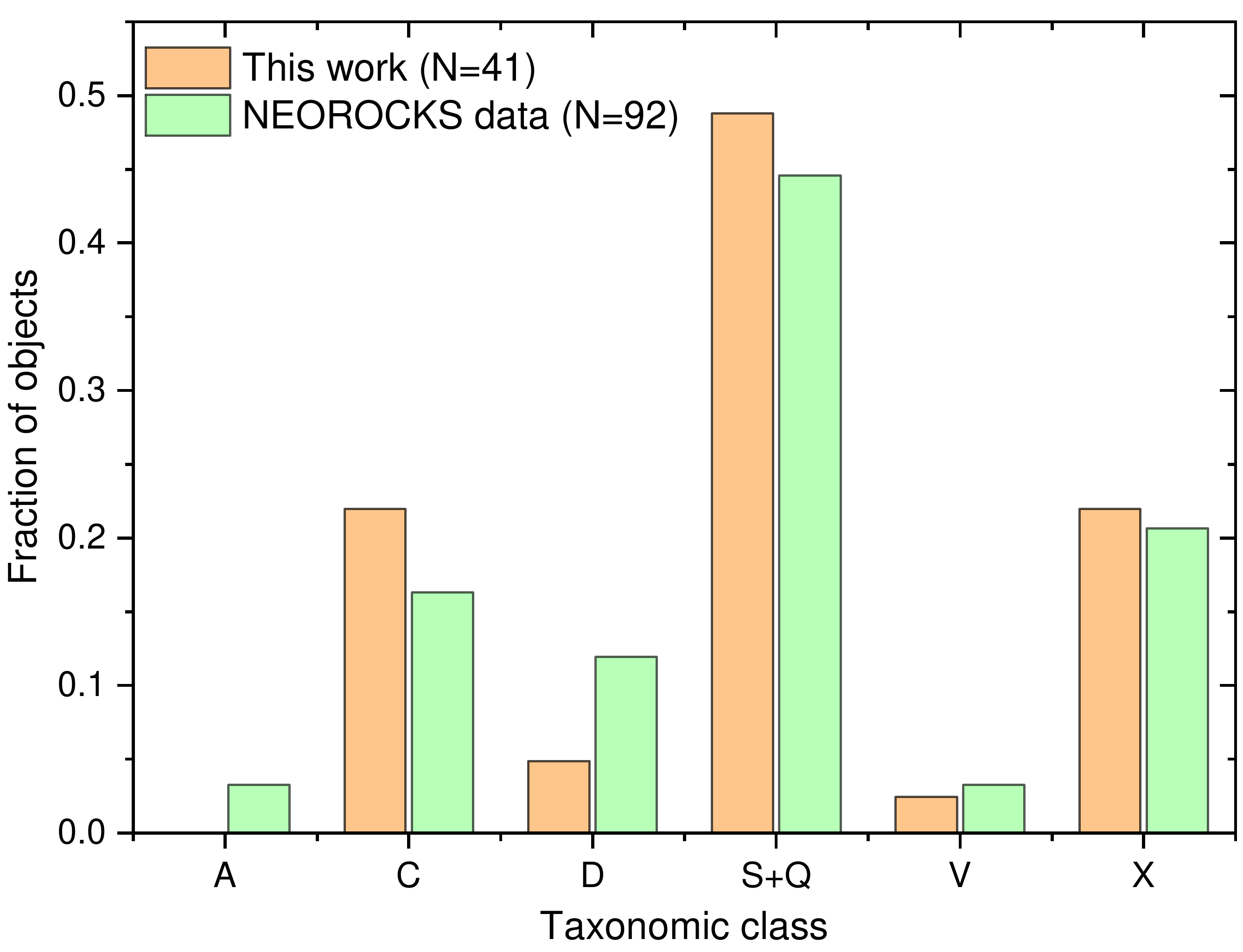}
    \caption{Distribution of the taxonomic classes obtained in this work and the overall distribution for the NEOROCKS survey data.}
    \label{hist}
\end{figure}

For the new data the taxonomic classes are distributed as follows: 20 objects are classified as S-complex, both C- and X-complexes have 9 objects, 2 objects are D-type, and 1 object is classified as V-type. No objects were classified as A-type.  
By adding new surface color data for 42 NEOs, the NEOROCKS dataset increased to 93 objects. 
Considering the results obtained by \citet{Hromakina2021}, in the extended sample of 92 objects (2022 GY2 not included) 46\% are classified into S-complex, 18\% into C-complex, 18\% into X-complex, 13\% as D-type, 3\% as A-type, and 3\% as V-type (Fig.~\ref{hist}). 

Nearly a half of NEOs in our dataset fall into the S-complex. This predominance of silicate S-complex asteroids among NEOs is consistent with existing surveys on NEOs. 
Such prevalence of S-type asteroids among NEOs can be affected by an observational bias towards objects with higher albedo. Thus, to obtain knowledge about the real compositional distribution, a debiasing technique should be applied. Given that the size of our sample is still relatively small, the debiasing of our data is beyond the scope of this work. 
We note that in the recent work by \citet{Marsset2022} the authors presented a debiased compositional distribution of NEOs. In particular, the portion of S-complex in their dataset drops from 66 to 42\% for the bias-corrected distribution, and the amount of primitive C-complex asteroids doubles and goes from about 20 to about 40\%. 

As a significant portion of our data consists of PHAs, we looked for possible differences in the compositional distribution of PHAs and non-PHAs, and found no significant difference in the overall distribution between these two groups. 
The lack of compositional difference between PHAs and the rest of the NEOs population was also reported by \citet{Perna2016}.

For further analysis, we split the classified objects in our dataset into groups of silicate objects, that includes S-complex, A-, and V-types, and carbonaceous objects, that consists of C-complex, D-type, and low-albedo X-types (also known as P-types in the Tholen taxonomy \citep{Tholen1984}). The remaining objects, namely X-complex asteroids with unknown albedo, are put into the miscellaneous group. This resulted in 47 objects in the silicate group, 28 objects in the carbonaceous group, and 17 objects in the miscellaneous group.

\begin{figure}
    \centering
    \includegraphics[width=0.95\columnwidth]{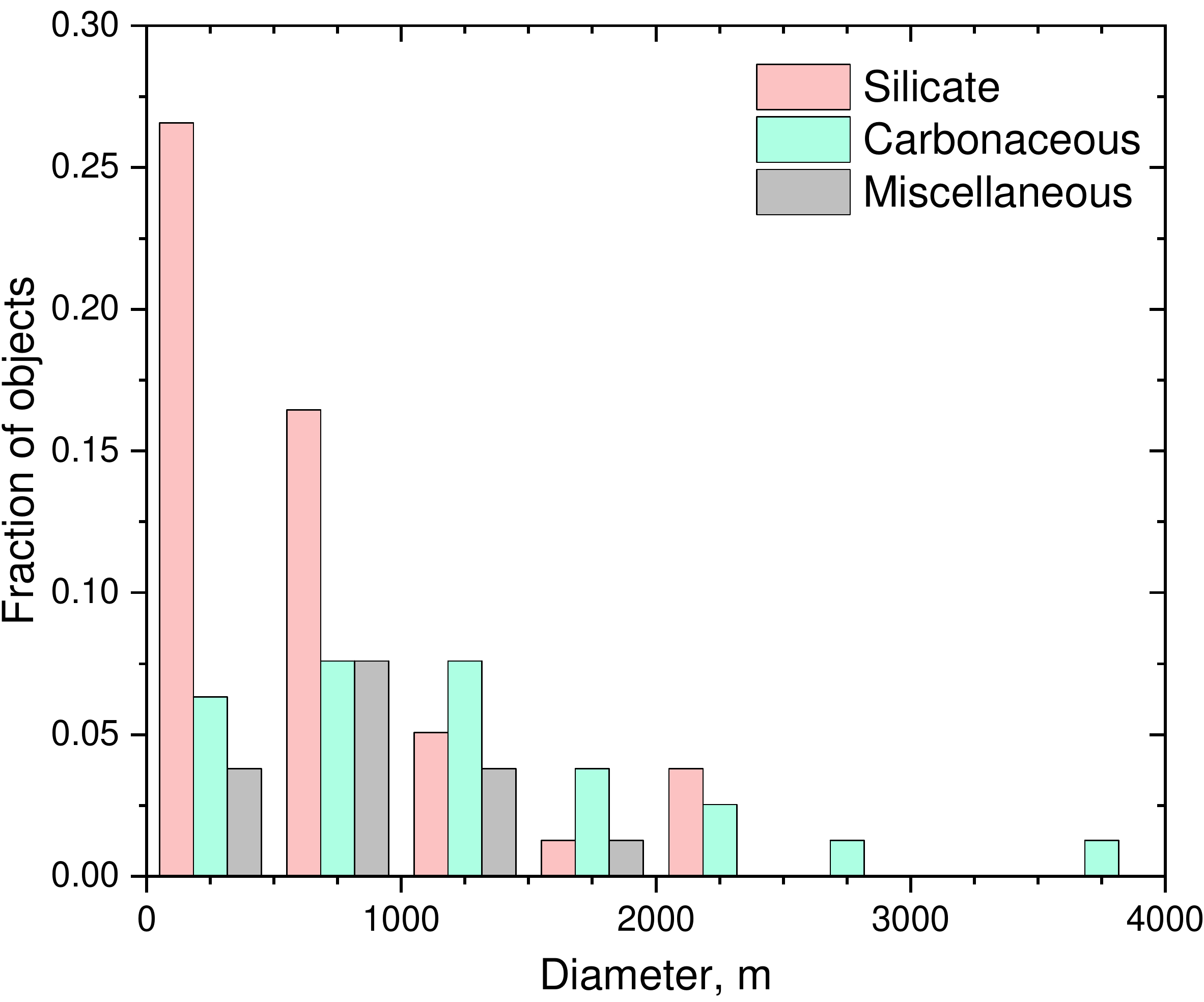}
    \caption{Size distribution of 92 NEOs in our dataset.}
    \label{D_hist}
\end{figure}

Figure~\ref{D_hist} shows diameter distribution of the silicate, carbonaceous, and miscellaneous objects in the NEOROCKS dataset.
The estimated sizes of the objects spread out from about 20 meters to almost 4~kilometers, with the majority of objects being in the D<1~km range. The number of carbonaceous objects in our dataset steadily decreases with size. Silicate objects are much more prevalent among smaller objects, but its number drops significantly for D>1~km. The relative ratio of carbonaceous to silicate objects with D<500~m is 28\% to 72\%, and for objects with D>500~m the ratio is 46 to 54\%. Such distribution is most probably caused by the bias towards higher albedo objects.

We looked into the possible differences in the orbital and physical parameters between the different groups of objects analysed in this work. In Table~\ref{table_mean_orb} we present the median values of the main orbital elements, absolute magnitude, as well as the estimated diameter for the whole dataset, groups of silicate and carbonaceous objects, and also for the PHAs and non-PHAs. The "dispersion" of the median values was calculated as a median absolute deviation (MAD).
Carbonaceous objects show slightly higher orbital inclinations and larger estimated diameters compared to the silicate ones. The Earth's MOID is lower for silicate objects, which in turn also suggests a generally lower MOID for smaller asteroids in our dataset because as we already established, the majority of objects with D<500~m belong to the silicate group.
PHAs in the dataset, as expected, have smaller $q$, $Q$, and $a$ values, as well as slightly smaller diameters than non-PHAs. Otherwise, we do not find any significant correlations between the orbital and physical parameters for the objects in our dataset. 

\begin{table*}
\small
    \centering
    \caption{Median values of the orbital and physical parameters for the whole dataset and different sub-groups discussed in this work. The uncertainty value was calculated as median absolute deviation.}
    \label{table_mean_orb}
        \begin{tabular}{lllllllll}
 
\hline\hline        
Group & $e$ & $i$, deg & $a$, au & $q$, au & $Q$, au & MOID, au & $H$ & $D$, m \\
\hline
All data (N=92) &0.49$\pm$0.08&10.38$\pm$6.24&1.78$\pm$0.50 &0.94$\pm$0.19&2.74$\pm$0.83 &0.06$\pm$0.05& 19.00$\pm$1.15& 572$\pm$290\\
PHAs (N=36)& 0.50$\pm$0.09&9.92$\pm$6.61& 1.58$\pm$0.50 & 0.78$\pm$0.18& 2.39$\pm$0.89& 0.02$\pm$0.01& 19.30$\pm$1.10 &540$\pm$268\\
non-PHAs (N=56)& 0.49$\pm$0.06& 11.38$\pm$6.50& 2.12$\pm$0.49 & 1.03$\pm$0.15&3.12$\pm$0.83&0.12$\pm$0.06 &18.85$\pm$1.15& 666$\pm$263\\
Silicate (N=47)& 0.48$\pm$0.04 &11.38$\pm$7.42 &1.95$\pm$0.39& 0.95$\pm$0.14& 2.93$\pm$0.71&0.05$\pm$0.03& 19.00$\pm$1.10 & 495$\pm$305\\
Carbonaceous (N=28)& 0.50$\pm$0.08 & 12.35$\pm$6.28 &1.65$\pm$0.53 &0.92$\pm$0.24& 2.50$\pm$0.84 &0.09$\pm$0.08 &19.00$\pm$1.30 & 808$\pm$382\\
Miscellaneous (N=17)& 0.54$\pm$0.09 &9.79$\pm$4.63 &1.69$\pm$0.60 &0.97$\pm$0.23 & 2.71$\pm$1.13  &0.05$\pm$0.04 & 18.7$\pm$1.10 & 603$\pm$269\\
    \hline
    \end{tabular}
\end{table*}

\begin{figure}
    \centering
    \includegraphics[width=0.95\columnwidth]{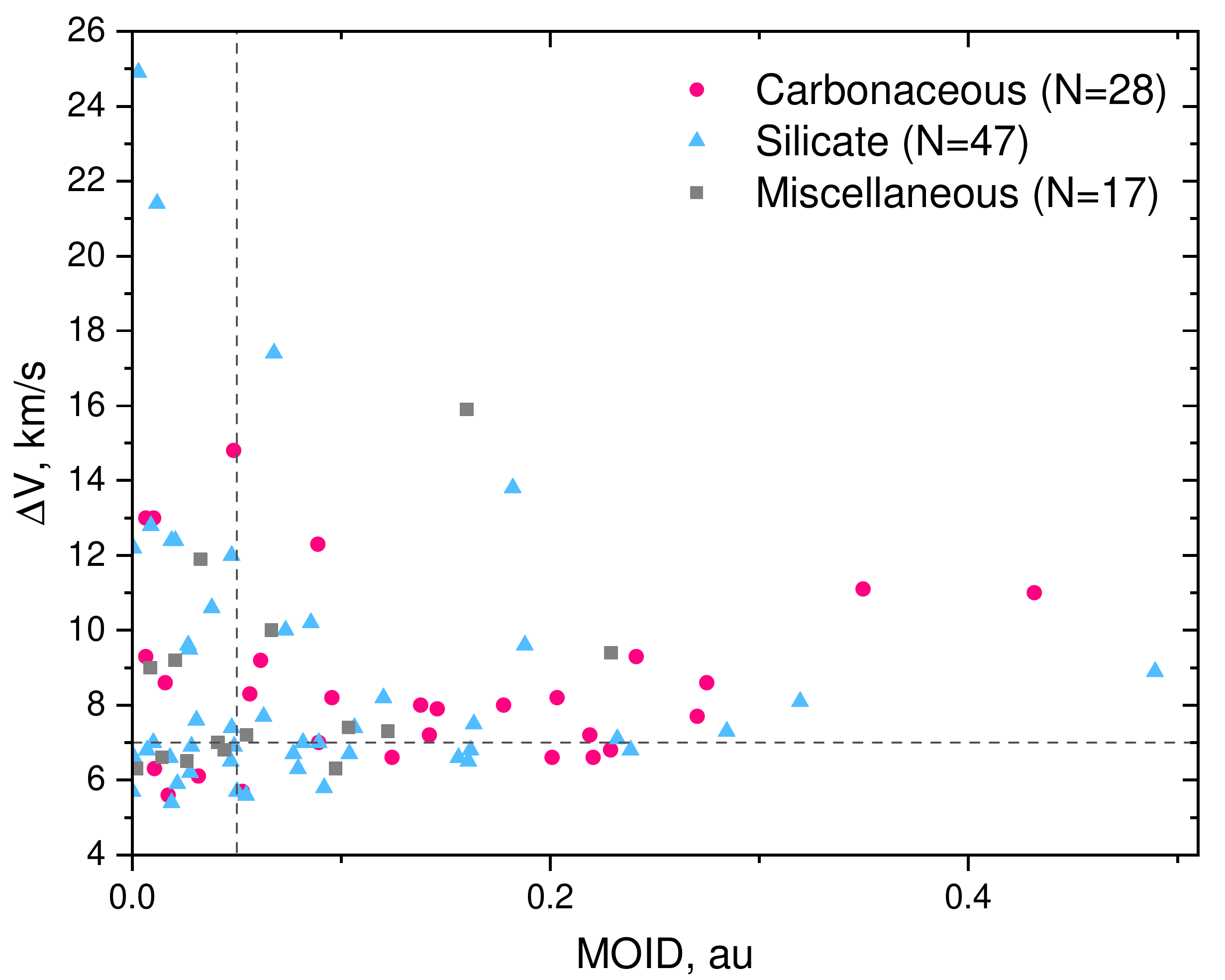}
    \caption{Earth's MOID vs. $\Delta$V value for the NEOs in our dataset. Vertical and horizontal dashed lines at MOID = 0.05~au and $\Delta$V = 7~km/s, respectively, separate objects that may be suitable as space mission targets.}
    \label{fig_orbital}
\end{figure}

As already mentioned, close approaches of NEOs to Earth raise a concern due to the possible impact as well as create a possibility to investigate these objects \textit{in situ} by a space mission.
In this regard, objects with smaller Earth's MOIDs (i.e. PHAs) are better candidates for space mission targets. Another parameter that should be taken into consideration is $\Delta$V, which is a measure of the impulse necessary for a spacecraft to reach the orbit of a target. According to \citet{Hinkle2015}, with current propulsion technology, only objects with $\Delta$V<7~km/s can be accessed by a spacecraft. Figure~\ref{fig_orbital} shows Earth's MOID and $\Delta$V values, that were taken from the MPC website, for the observed objects. In total, 14 objects have both MOID<0.05~au and $\Delta$V<7~km/s (Table~\ref{table_deltaV}).
The majority of them (6 out of 10) belong to the silicate group, 3 objects are from the miscellaneous group and only one is from the carbonaceous group. 
Of particular interest are D-type and rare A-type asteroids, because they have not been visited by space missions yet \citep{Barucci2018}. In this regard, we can mention two objects from our data: D-type NEO (163014) 2001 UA5 and A-type NEO 2017 SE19. In Table~\ref{table_deltaV} we also show the values of the parameter U, which corresponds to the orbit uncertainty, where 0 corresponds to a very small uncertainty and 9 indicates a very large uncertainty.  

\begin{table}
\small
    \centering
    \caption{List of NEOs that could be potential space mission targets based on their low MOID and low $\Delta$V.}
    \label{table_deltaV}
        \begin{tabular}{llll}
\hline\hline        
Object & Taxon & $\Delta$V, km/s & U$^*$\\
    \hline
(12923)  Zephyr&S-complex&5.9&  0  \\
(52768) 1998 OR2&X-complex&6.6&  0  \\
(140158) 2001 SX169&X-complex&6.5& 0 \\
(332446) 2008 AF4&X-complex&6.3& 0 \\
(388945) 2008 TZ3&Cg/C-complex&5.6& 0\\
(163014) 2001 UA5&D-type&6.1& 0 \\
(613403) 2006 GB&C-complex&6.4&0\\
2009 CC3&S-complex&6.9&0\\
2010 TV149&X-complex&6.8&0\\
2017 SE19&A-type&6.5&  0 \\
2002 TP69&S-complex&5.7& 0 \\
2020 RO6&S-complex&5.4& 0 \\
2020 YQ3&S-complex&6.8& 0 \\
2022 BH3&S-complex&6.6& 6 \\
    \hline
    \end{tabular}
\begin{flushleft}
$^*$Orbit uncertainty parameter taken from the MPC website.
    \end{flushleft}  
\end{table}

\section{Summary and conclusions}
\label{concl}

In this work we present new observational data for 42 small NEOs obtained within the NEOROCKS project. For these asteroids we computed surface colors, and estimated their taxonomic class, absolute magnitudes, and diameters. 
Based on the B-V, V-R, and V-I surface colors and considering only the main taxonomic classes, most of the observed objects (20 NEOs) were classified into the S-complex, 9 as C-complex, 9 as X-complex, 2 as D-type, and 1 as V-type.

By adding these new observational data we expanded our dataset to 93 objects. 
The compositional distribution of the updated dataset is as follows: most of the objects (46\%) were classified into the S-complex, 18\% into the X-complex, 18\% into C-complex, and 13\% as D-type. The remaining small number of objects were classified as A-types (3\%) and V-types (3\%). The calculated mean colors for the taxonomic classes considered in this work are in agreement with the literature values \citep{DeMeo2009,Ieva2018,Lin2018}. PHAs do not show a significant difference in the compositional distribution from the rest of the population.

Asteroids in our dataset with sizes D<500~m predominantly belong to the silicate group. This result is due to the observational bias towards higher albedo objects rather than an intrinsic compositional difference.

The group of silicate objects shows lower Earth MOID compared to the group of low-albedo carbonaceous objects. Silicate objects also show a slightly smaller median inclination, but more data is needed to confirm a possible difference. 

Finally, we selected objects that could be considered as space mission targets based on their MOID and $\Delta$V values. Among 10 objects in our dataset that satisfy both criteria, we particularly note primitive D-type asteroids (163014) 2001 UA5 and rare A-type asteroid 2017 SE19.

The NEOROCKS project is still ongoing and new observational runs are scheduled. In the final article, which will be done at the end of the project, we will summarize the obtained results on the surface colors of NEOs and perform a global statistical analysis that will include available literature data.

\section*{Acknowledgements}

We acknowledge funding from the European Union’s Horizon 2020 research and innovation programme under grant agreement No. 870403. The article is based on observations made at Observatoire de Haute Provence (CNRS), France.

%%%%%%%%%%%%%%%%%%%%%%%%%%%%%%%%%%%%%%%%%%%%%%%%%%
\section*{Data Availability}

The data underlying this article will be shared on reasonable request to the corresponding author.



\bibliographystyle{mnras}
\bibliography{mnras_NEOROCKS} % if your bibtex file is called example.bib

%\begin{thebibliography}{99}
%\end{thebibliography}

%%%%%%%%%%%%%%%%%%%%%%%%%%%%%%%%%%%%%%%%%%%%%%%%%%

%%%%%%%%%%%%%%%%% APPENDICES %%%%%%%%%%%%%%%%%%%%%

\appendix

\section{Orbital elements of the observed objects}
\label{annex-table}

\begin{table*}
\small
    \centering
    \caption{List of observed objects and corresponding orbital elements taken from the MPC website.}
    \label{table1}
        \begin{tabular}{llllllll}
\hline
\hline
&Object&Type & e&i, deg&a, au&q, au & Tj\\
\hline
1&(12923) Zephyr&Apollo, PHA& 0.4922&5.305&1.9620&0.9962&3.716 \\
2&(87024) 2000 JS66&Apollo &0.1899&14.431&1.1967&0.9694&5.260   \\
3&(89958) 2002 LY45&Apollo, PHA &  0.8862&9.908&1.6417&0.1867&3.682\\
4&(138971) 2001 CB21&Apollo, PHA&  0.3333&7.899&1.0343&0.6895&5.863\\
5&(140158) 2001 SX169&Apollo, PHA & 0.4609&2.518&1.3472&0.7262&4.764\\
6&(143649) 2003 QQ47&Apollo, PHA&  0.1870&62.102&1.0850&0.8820&5.214\\
7&(162913) 2001 MT18&  Apollo&0.5196&8.647&1.2709&0.6105&4.929\\
8&(163692) 2003 CY18&Apollo& 0.4106&7.199&1.5266&0.8998&4.389\\
9&(318160) 2004 QZ2&Amor&   0.4929&0.9691&2.261&1.1465&3.448\\
9&(363027) 1998 ST27&Aten, PHA&  0.5300&21.062&0.8190&0.3850&6.978\\
10&(363599) 2004 FG11&Apollo, PHA&0.7238&3.1269&1.5872&0.4384&4.039\\
12&(366746) 2004 LJ&Apollo, PHA &  0.4616&18.282&1.0865&0.5850&5.559\\
13&(388945) 2008 TZ3&Apollo, PHA&0.3907&8.7176&1.5900&0.9689&4.278\\
14&(374855) 2006 VQ13&Apollo, PHA&0.4457&16.698&1.0993&0.6093&5.521\\
15&(410195) 2007 RT147&Amor &    0.4668&3.831&2.2912&1.2216&3.442\\
16&(415029) 2011 UL21&Apollo, PHA &  0.6528&34.851&2.1228&0.7369&3.245\\
17&(450263) 2003 WD158&Apollo, PHA &  0.4091&16.726&1.4264&0.8428&4.563\\
18&(475665) 2006 VY13&Amor&0.6137&4.673&4.6727&4.6727&2.960\\
19&(491567) 2012 RG3&Apollo&  0.5286&15.918&1.5744&0.7422&4.203\\
20&(495615) 2015 PQ291&Aten&   0.2374&34.602&0.9825&0.7493&5.991\\
21&(506459) 2002 AL14&Apollo&  0.1261&22.998&1.0376&0.9067&5.830\\
22&(516396) 2000 WY28&Amor&  0.2894&19.447&1.6384&1.1642&4.189\\
23&(613291) 2005 YX128&Apollo&  0.7347&4.476&2.1744&0.5767&3.267\\
24&(613403) 2006 GB&Aten, PHA&0.1793&10.061&0.9590&0.7871&6.257\\
25&2002 TP69&Amor& 0.4695&1.974&1.9475&1.0331&3.751\\
26&2009 CC3&Amor, PHA&0.5373&11.372&2.2006&1.0182&3.440\\
27&2010 TV149&Apollo, PHA&0.5188&9.792&1.6989&9.7926&4.025\\
28&2010 WQ7&Amor&   0.5567&13.610&2.3699&1.0504&3.285\\
29&2011 YQ10&Amor&   0.5216&2.397&2.2418&1.0724&3.440\\
30&2017 UW42&Amor &   0.4842&6.670&2.4560&1.2667&3.313\\
31&2018 CW13&Apollo&   0.7051&28.918&1.6108&0.4750&3.921\\
32&2021 JQ24&Amor&  0.6695&5.993&3.2256&1.0659&2.776\\
33&2021 LN3&Apollo&  0.5662&28.80&2.2776&0.9881&3.241\\
34&2021 SR41&Apollo&0.6502&21.935&2.3796&0.8324&3.140\\
35&2021 VM25&Apollo&0.3913&9.162&1.6474&1.0028&4.181\\ 
36&2021 WX3&Amor&0.5144&7.853&2.1988&1.0677&3.471\\
37&2022 BH3&Apollo&  0.5972&3.943&2.1629&0.8711&3.437\\
38&2022 BQ3&Amor&0.6027&15.088&2.8279&1.1237&2.976\\
39&2022 BK&Amor &  0.5801&10.713&2.4666&1.0357&3.212\\
40&2022 DC5&Apollo&0.2061&3.925&1.0142&0.8052&2.976\\
41&2022 GC1&Apollo&0.4006&13.331&1.1695&0.7009&5.294\\
42&2022 GY2&Apollo, PHA&0.8647&6.199&2.1673&0.2933&3.049\\
    \hline
    \end{tabular}

\end{table*}

\section{Individual reflectance spectra of the targets}
\label{annex-figure}

\begin{figure*}
    \centering
    \includegraphics[width=\textwidth]{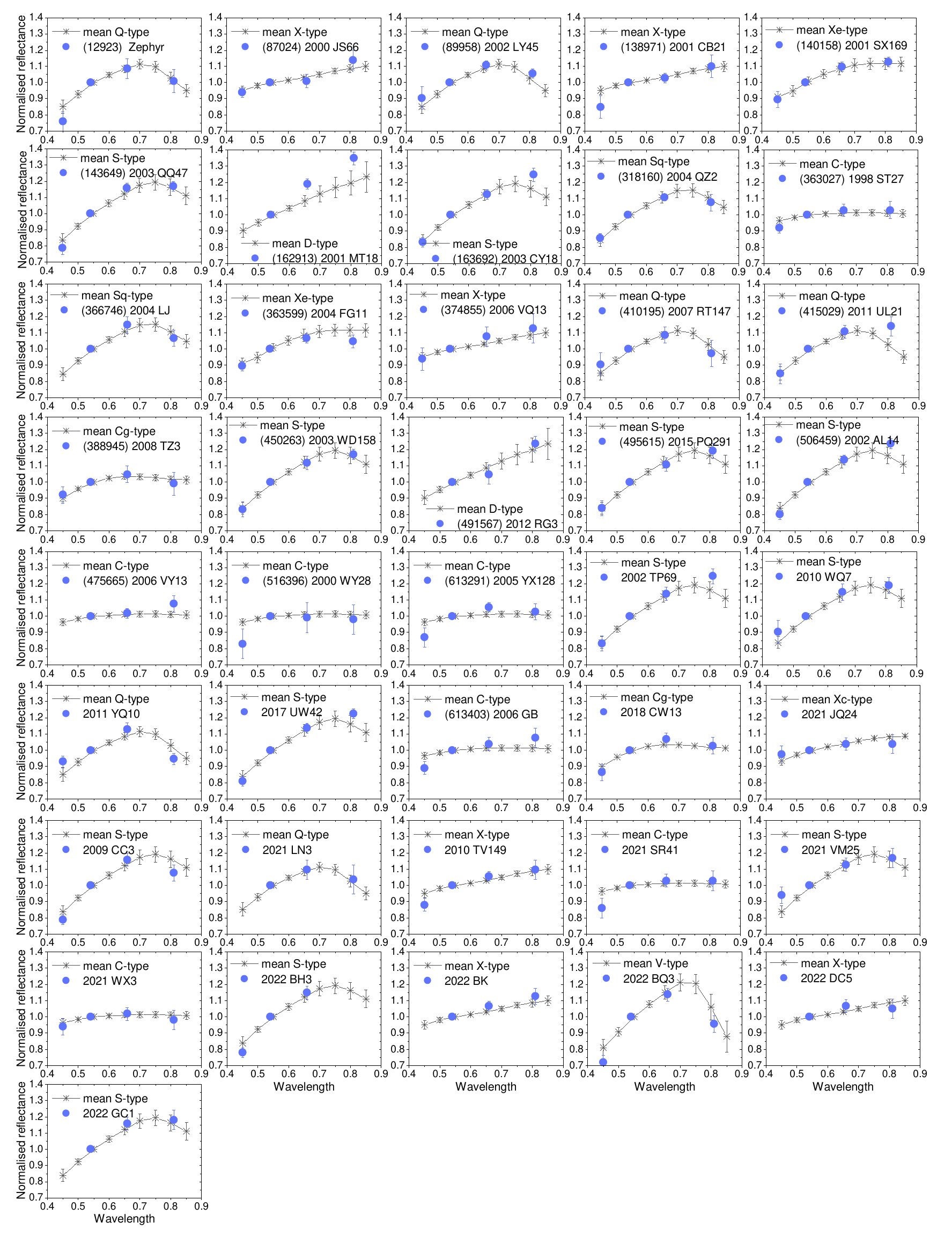}
    \caption{
    Individual reflectance spectra of the objects observed in this work (blue dots) together with the mean spectra of the most fitting taxonomic class from \citet{DeMeo2009} (black asterisks). The spectra are normalized at 0.54~$\mu$m.}
    \label{indiv_plots}
\end{figure*}

%%%%%%%%%%%%%%%%%%%%%%%%%%%%%%%%%%%%%%%%%%%%%%%%%%

% Don't change these lines
\bsp	% typesetting comment
\label{lastpage}
\end{document}